\documentstyle[prl,aps,multicol,epsf,epsfig]{revtex}
\def\S{{\bf S}}
\begin{document}
\title{Superconductivity in CoO$_2$ Layers and the  Resonating Valence Bond Mean Field Theory of the Triangular 
       Lattice t-J Model}
\author{Brijesh Kumar and B. Sriram Shastry }
\address{Department of Physics, Indian Institute of Science, Bangalore 560012, India.}
\maketitle
\begin{abstract}
Motivated by the recent discovery of superconductivity in two dimensional CoO$_2$ layers,
 we present some possibly useful results of the RVB mean field theory applied to the 
triangular lattice. An interesting  time reversal breaking superconducting state arises from  strongly
  frustrated interactions. Away from half filling, the order parameter is found to be complex, and yields a fully gapped 
quasiparticle spectrum. The sign of the hopping plays a crucial role in the analysis, 
and we find that superconductivity is as fragile for one sign as it is robust for the other. Na$_x$CoO$_2\cdot y$H$_2$O 
is argued to belong to the robust case, by comparing the LDA fermi surface with an effective tight binding model. 
The high frequency Hall constant in this system is potentially interesting, since it is pointed out to increase linearly 
with temperature without saturation for T $>$ T$_{degeneracy}$.

\vspace{5mm}
\end{abstract}
\begin{multicols}{2}
\section{Introduction}
The recent discovery of superconductivity at low temperatures in CoO$_2$ layered compounds~\cite{takada}
is an exciting event, since it may be the long sought {\it low temperature} RVB superconductor, on a lattice which 
was at the basis of Anderson's original ideas on a possible quantum spin liquid state~\cite{anderson_rvb}. 
Although the spin-1/2 triangular lattice appears to have better states with 
three sublattice magnetic order, it is possible that the RVB state is attained for sufficiently high doping, and 
it seems to be both useful and worthwhile to explicitly state the  detailed results of the RVB ideas applied to this lattice so
as to serve as a reference point for further experiments that are surely forthcoming shortly.

 Another reason why the triangular lattice is important is that there exists a very complete and highly nontrivial set 
of ideas having their origin in the Fractional Quantum Hall physics that have been theoretically applied to  the 
triangular lattice by Kalmeyer and Laughlin~\cite{kalmeyer_laughlin}. Their picture of a m=2 FQHE state of 
interacting hardcore bosons (viz the spin-1/2 particles ), leads to anyonic particles on doping
and to a Meissner like time reversal breaking state. Such a state can be alternately viewed in the language of 
the flux phases, where Anderson, Shastry and Hristopulous~\cite{ASH} and also Reference~\cite{lee_feng}, 
showed its equivalence to a flux $\pi/2$ per triangle state . 
Systematics of the doping dependence of the optimum flux have not apparently been done, and presumably that is another
interesting area to pursue in the present context.

Earlier  weak coupling results on superconductivity in the triangular lattice \cite{vojta_dagotto} were motivated by
experiments on organic superconductors. A recent high temperature expansion study of the t-J model on the triangular
lattice estimates the  total entropy and magnetic susceptibility as  a function of temperature and doping \cite{koretsune_ogata}.

In the present work, we perform what seems to be a consistent and simple version of RVB theory, one which yields 
d-wave order for the square lattice~\cite{kotliar} and also for the other recently interesting case of 
SrCu$_2$(BO$_3$)$_2$~\cite{bss_bk}.  Earlier calculations have been done by other
authors  in a similar spirit to ours,
Reference~\cite{lee_feng} is confined to half filling,  and a recent preprint by Baskaran~\cite{baskaran} 
makes some qualitative points that are common to our
calculations.  Within this version, we evaluate the case of positive as well as negative hoppings 
since these are so very different in their physical content. Using the particle-hole transformation for fermions, we 
may define two broad cases of interest:
\begin{itemize}
 \item Case A: Here we have either 
\subitem (i) $t>0$ and electron doping, or 
\subitem(ii) $t<0$ with hole doping, and
\item Case B: where we have either
\subitem(i) $t>0$ and hole doping, or
\subitem(ii) $t<0$ and electron doping.
\end{itemize}
Here we note that the hamiltonian of the t-J model is written in the standard form : 
\begin{equation}
\mbox{H}=-t\sum_{<i,j>,\sigma}{\cal P}c^\dag_{i\sigma}c_{j\sigma}{\cal P} + 
          \mbox{J}\sum_{<i,j>}(\S_i\cdot\S_j-\frac{n_in_j}{4})
\label{tj_ham}
\end{equation}
where ${\cal P}$ stands for the Gutzwiller  projection due to large U, and the summation is over nearest neighbors. The notation of hole and electron filling is {\it relative 
to half filling} in the effective one band model for this system, and as usual, $\delta=|1-n|$, where $n$ is the
electron concentration. This model incorporates both electronic frustration and spin frustration through the kinetic and
the exchange energies.  The former  favors Nagaoka  ferromagnetism for Case A (see later),  while the frustrated
antiferromagnetic exchange J competes against it for both cases. The use of an antiferromagnetic exchange is motivated by
the observed Curie Weiss susceptibility with  a negative Curie Weiss temperature\cite{ray}, and though surprising in the
event of a $98^0$ Co-O-Co bond angle, is not without precedent, since CuGeO$_3$ with the same bond angle has also an AFM J 
($\sim 120^0$K)\cite{nishi_fujita}.

\begin{figure}
\epsfxsize=7.5cm
\epsffile{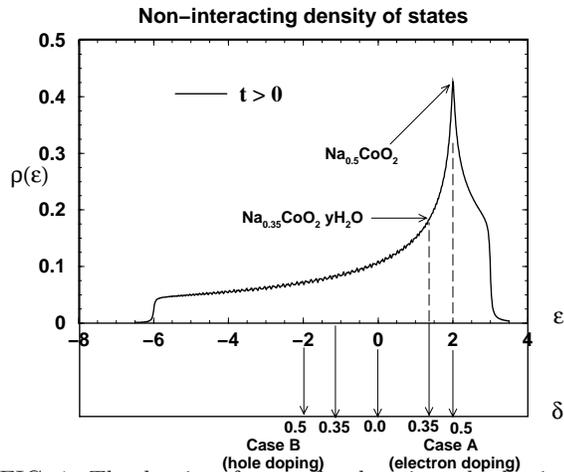}
\caption{The density of states for the triangular lattice tight binding hamiltonian, and locations of various systems.}
\label{dos_ni}
\end{figure}

 A first step in the direction of identifying an effective one band model is in the work of 
Singh~\cite{dsingh}, whose LDA calculation of the band structure shows that the fermi energy for the case of 
NaCo$_2$O$_4$ is in a tight binding like set of states of t$_{2g}$ symmetry, in the close proximity of and slightly 
above a sharp  peak in the electronic density of state. We interpret this as an example of Case A(i) above, since
 the fermi surface for triangular lattice tight binding band structure, $\epsilon({\bf k}) = 
-2t(\cos(k_x)+2\cos(k_x/2)\cos(\sqrt{3}k_y/2)$, gives a density of states (DOS) (see Figure~\ref{dos_ni}) 
with a prominent peak near the fermi level for Na$_{0.5}$CoO$_2$ as well as the extremity of the band, and 
to the extent that the low energy structures are irrelevant, this matches the LDA DOS\cite{comment}. We return to discuss
this issue  in the summary section.
The fiduciary Mott insulating state from where the hole/electron doping is measured, is  the case of pure CoO$_2$ on a 
triangular lattice.  This particular system has apparently   not been realized experimentally  so far due to a lattice structure change.

\section{A few remarks on ferromagnetism and the Hall constant in the triangular lattice}
Before describing the RVB calculation, we recount a few remarkable features of the triangular lattice  
Hubbard and t-J  model physics, that may be useful in future studies.

\subsection{Ferromagnetism}
Singh has noted that the LDA calculations of NaCo$_2$O$_4$ show an instability of the 
paramagnetic state towards a ferromagnetic state.
Indeed this is exactly what one expects from the Nagaoka physics on the triangular lattice as shown by Shastry, 
Krishnamurthy and Anderson~\cite{bss_hrk_pwa}, who pointed out that while Case B above is highly detrimental to 
ferromagnetism in the  infinite U limit, Case A highly favors the ferromagnetic state. This follows from a stability 
analysis of the low energy excitations of the state.
 Ferromagnetism is the fate of the  Case A, t-J model at ${\rm J}=0$ for essentially all fillings. At low dopings,
turning on a  sufficiently strong antiferromagnetic J   removes the Nagaoka instability,
for some range of hole doping. We argue below that in this very doping range,
the RVB superconducting phase emerges instead.
At higher doping, i.e.  in the high electron  density limit of Case A(i),   J becomes 
irrelevant.   For an almost filled band, with the fermi energy near a peak in the DOS (as in transition metal ferromagnetism) the work of Kanamori  and Galitskii predicts ferromagnetism. Thus notwithstanding the results 
of the RVB state, we must expect metallic ferromagnetism in the case when the electron doping is high in Case A(i).
The observation of ferromagnetism in Na$_{0.75}$CoO$_2$\cite{motohashi} is consistent with these arguments, while 
being enigmatic in that the high temperature susceptibility shows a negative Curie Weiss temperature. 
\subsection{High Frequency Hall Effect} 
 A fascinating property of the triangular lattice was noted in Reference~\cite{bss_bs_rrps}.   The high 
frequency Hall constant $R_H^*$  is amenable to a lattice walk expansion.  It can be expressed
in terms of loops encircling a flux, and  manages to capture the Mott Hubbard aspect of the
problem, such as a vanishing  ``effective Hall carrier density'' near half filling on the square lattice.   The resulting high frequency
 Hall constant,  in Mott Hubbard systems is {\em not } a measure of
carrier density, unlike in simple conductors, but encodes complicated correlations of the underlying system.
The topology of the  lattice plays a critical role in determining this object, since it
depends upon the length of the closed loops, and the triangular lattice was noted to be exceptional in having
the smallest length  of  a closed loop,with an odd number of steps  namely 3, leading to a very different
behaviour from say the square lattice. 
  A  calculation  for the triangular lattice  in the case of  hole doping yields
\begin{equation}
R_H^* =   - \frac{v}{8 |e|} \frac{k_B T}{ t}\frac{1+\delta}{\delta (1- \delta)}  \;\mbox{(Cases A(ii) or B(i))}.
\label{hall_coefficient_hole}
\end{equation}
 Here $|e|$ is the magnitude of the
 electronic charge  and $v$ is the physical (three dimensional) unit cell 
volume containing one  cobalt ion, which from Reference~\cite{takada} may be estimated to be $67.71\times 10^{-24}$
cm$^3$~\cite{note1}. We remark that this result is computed in the case of hole doping and either sign of 
hopping ``$t$'', i.e. Case A(ii) or B(i), and in comparing with electron doping, A(i) or B(ii), one must use the usual
 rules for particle-hole transformation $t \rightarrow - t$ as well as $\delta = |1-n|$. This leads to the following
expression for $R_H^*$ in the case of electron doping,
\begin{equation}
R_H^* =   \frac{v}{8 |e|} \frac{k_B T}{ t}\frac{1+\delta}{\delta (1- \delta)}\; \mbox{(Cases A(i) or B(ii))}.
\label{hall_coefficient_electron}
\end{equation}
The expressions in Equations~\ref{hall_coefficient_hole} and~\ref{hall_coefficient_electron} are valid when
temperature, T$>$T$_{degeneracy} \sim \, |t|$. Since T$_{degeneracy}$ seems so low in these 
systems, as evinced by the strong Curie-Weiss behavior, this  result, so  remarkable in its absence of saturation in 
temperature,  seems worthwhile to check experimentally. This could also be used to
experimentally determine the magnitude as well as sign of ``$t$'' for an effective one band system.
The distinction between the transport and the high frequency Hall constants is argued to be 
through a weakly frequency and temperature dependent self energy, and hence it is possible that the transport 
measurements are also anomalous in the same sense\cite{note2}. 

\section {The RVB calculation and its detailed predictions} 
\subsection{Mean field equations}
The `J' term in Equation~\ref{tj_ham} can be re-written as \( -\mbox{J}\sum_{i,j} \mbox{b}^\dag_{ij}\mbox{b}_{ij} \),
where the bond operator, \(\mbox{b}^\dag_{ij} =
(c^\dag_{i\uparrow}c^\dag_{j\downarrow}-c^\dag_{i\downarrow}c^\dag_{j\uparrow})/\sqrt{2}\), is a singlet pair
creation operator acting on a pair of sites $i$ and $j$. The RVB mean field calculations are carried out by defining 
a complex order parameter, \(\Delta_{ij} = \langle b_{ij}\rangle\). On the triangular lattice, we have three different
nearest neighbor bonds, one along say x-direction and the other two at an angle of $\pi/3$ and $2\pi/3$ with x-axis.
We consider a simple situation where the $|\Delta_{ij}| = \Delta$ along all bonds, but three different phases are allowed 
along three kinds of bonds. Since one of the phases can be gauged away, only two (relative)
phases are sufficient. We assign zero phase along the x-direction, $\theta$ and $\phi$ along the $\pi/3$ and $2\pi/3$
directions, respectively. In momentum space, this choice of mean field order parameter leads to the following k-space
function, D({\bf k}), which carries the order parameter symmetry information.
\begin{eqnarray}
\mbox{D}({\bf k}) & = & \cos(k_x) + e^{i\theta}\cos( k_x/2 + \sqrt{3} k_y/2) \nonumber\\
                  &   & \hspace{1.6cm} +~e^{i\phi}\cos( k_x/2 - \sqrt{3} k_y/2)
\label{dee_k}
\end{eqnarray}

The effect of projection of double occupancy on hopping is accounted for by a simple approximation where $t$ is
replaced by $t\delta$, with $\delta$ being the hole concentration. 
Our calculations have been done only for hole doping case, but for both $t>0$ as well as $t<0$ 
(Cases  B(i) and  A(ii)). The case of electron doping can easily be related to these calculations by
particle-hole transformation, as described earlier.

We get two mean field equations, one for $\Delta$ and the other for $\mu$, the chemical potential. These are :
\begin{eqnarray}
\Delta = \frac{1}{6\mbox{J}L}\sum_{\bf k}\frac{\partial\mbox{E}({\bf k})}{\partial\Delta}
         \tanh(\frac{\beta\mbox{E}({\bf k})}{2}) \label{eq_Delta} \\
\delta = -\frac{1}{L}\sum_{\bf k}\frac{\partial\mbox{E}({\bf k})}{\partial\mu}\tanh(\frac{\beta\mbox{E}({\bf
         k})}{2}) \label{eq_mu}
\end{eqnarray}
Here, \(\mbox{E}({\bf k}) = \sqrt{(\epsilon({\bf k})-\mu)^2+2\mbox{J}^2\Delta^2|D({\bf k})|^2}\) , D({\bf k}) is given
in Equation~\ref{dee_k}.

\subsection{ Order parameter and Quasiparticle spectrum}
We solve Equations~\ref{eq_Delta} and~\ref{eq_mu} self-consistently for given values of $\delta$ and $t$, and for 
different choices of $\theta$ and $\phi$. All energies are measured in units of the exchange coupling, J. 

First, we
perform  the computation at T=0, and find the values of $\theta$ and $\phi$, at different hole concentrations, 
for which the ground state energy is minimum. This fixes the symmetry of the mean field order parameter. At $\delta=0$, 
 the ground state energy is lowest for $(\theta,\phi)=(0,\pm\pi/2)$ and $(\theta,\phi)=(\pm\pi/2,0)$, in contrast to 
the case of $(\theta,\phi)=(2\pi/3,4\pi/3)$ of Reference~\cite{lee_feng} which has slightly higher energy. 
At half filling, the ground state has a lower symmetry of cubic type rather than the six-fold rotational symmetry of 
the triangular lattice.
\epsfxsize=7.5cm
\epsffile{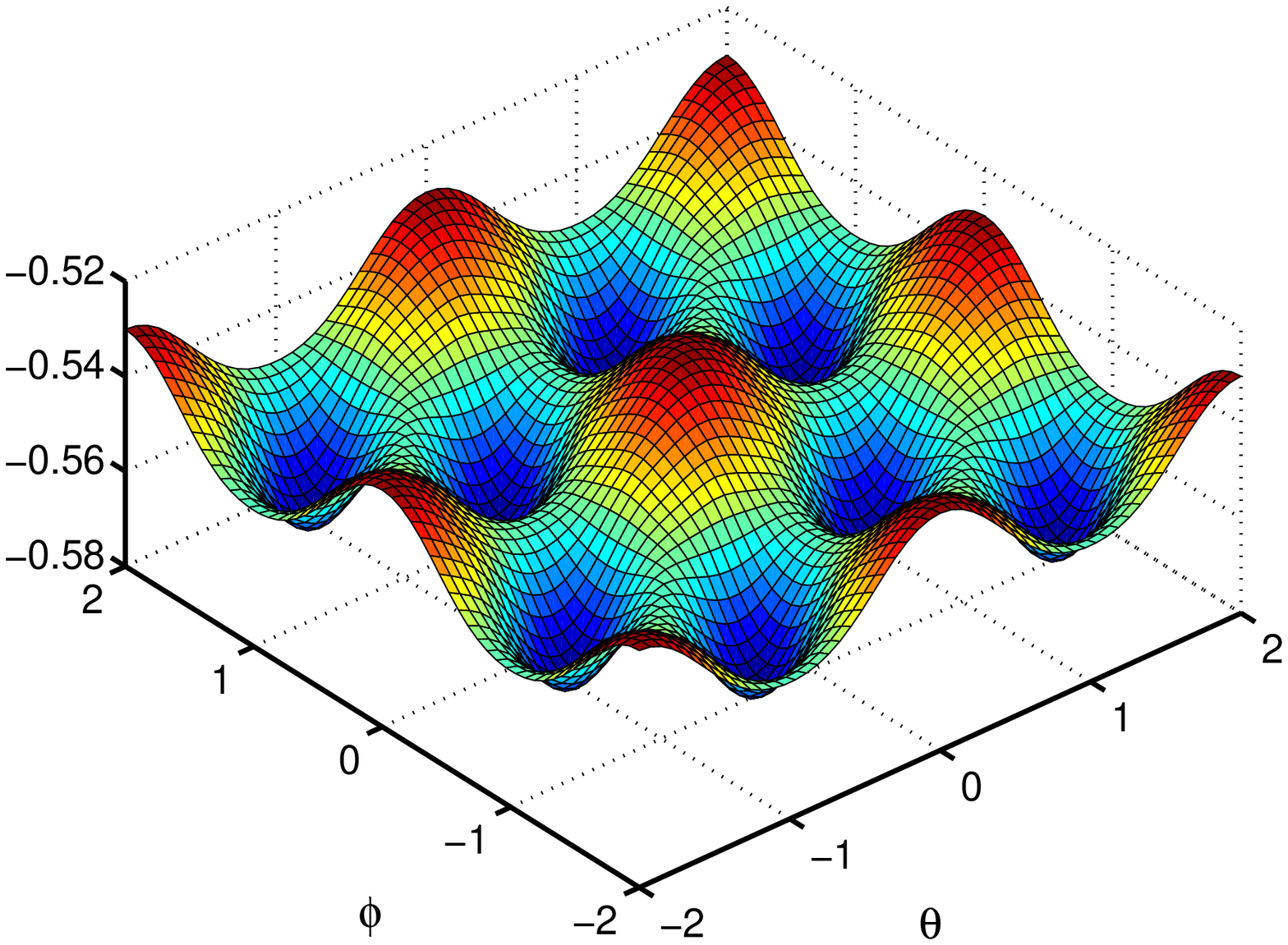}
\begin{figure}[h]
\caption{The mean field energy density surface  as a function of the internal phases, $\theta$ and $\phi$ (in units of
         $\pi$) of the order parameter. The minima occur at six corner points of the Brillouin zone.}
\label{phase_min}
\end{figure}
Away from 
half filling, the ground state energy is lowest at $(\theta,\phi)=(2\pi/3,4\pi/3)$,$(4\pi/3,2\pi/3)$ and 
$(2\pi/3,-2\pi/3)$. Three other phase-points, which are related to these via inversion with respect to origin, are
 equivalent, and together  these six minimum energy phase-points reflect the symmetry of the Brillouin zone. Figure~\ref{phase_min} shows the mean
field energy density plotted as a function of $\theta$ and $\phi$.
In rest of the calculations, we just choose one of these points to  perform the computation.

\begin{figure}
\epsfxsize=7.5cm
\epsffile{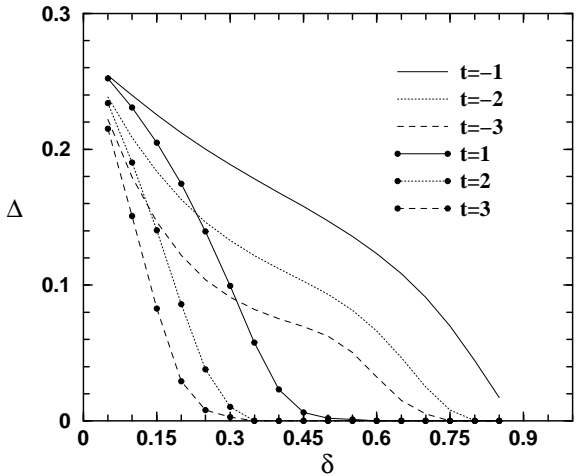}
\caption{The mean field $\Delta$ as a function of the hole concentration, $\delta$, for both positive and negative $t$ in units of J.}
\label{dD}
\end{figure}

At $\theta=2\pi/3$ and $\phi=4\pi/3$, for various values of $t$, the mean field $\Delta$ is computed as a function of
hole concentration, as shown in  Figure~\ref{dD}.  $\Delta$ decreases  rapidly for $t>0$ rather than for 
$t<0$, as the hole concentration is increased.  The larger magnitude of $\Delta$ as well as  the  greater doping  range
 suggests that Case A, that is `` $t<0$ and hole doping '' or 
`` $t>0$ and electron doping '', presents a  more robust case for the RVB state of superconductivity, as compared to the Case B.
The results of this low temperature mean field theory should be contrasted to those of the high temperature series 
results  of Ref(\cite{koretsune_ogata}), who interpret their extrapolated  results to imply lower entropy
 at low temperatures ($kT< |t|$)
for Case B, and hence perhaps greater tendency towards some (non ferromagnetic) ordering.  Further work needs to be done to
reconcile these findings.

Next we present the quasi-particle dispersion and the quasi-particle density of states from 
our mean field calculations. These calculation are  done for a
physically relevant value of $\delta \sim 0.35$ at which superconductivity is observed in Na$_x$CoO$_2\cdot y$H$_2$O~\cite{takada}.
The cobalt oxide layer alone acts as a half filled system in the effective one
band picture, therefore the carrier concentration in excess to the half filled case is just $x$. 

\begin{figure}
\epsfxsize=7.5cm
\epsffile{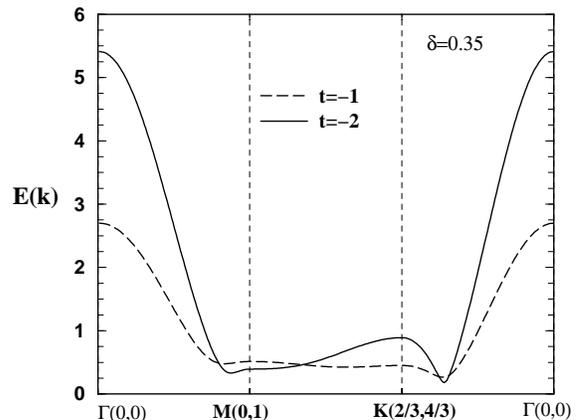}
\caption{The quasi-particle dispersion, E({\bf k}).
         The coordinates of $\Gamma$, M and K symmetry points in the Brillouin zone are given in units of $\pi$. 
         The coordinates ($k_1$,$k_2$) given here are such that $k_1 = k_x$ and $k_2 = (k_x + \sqrt{3}k_y)/2$, 
         where $k_x$ and $k_y$ are the usual k-space variables along $x$ and $y$ directions.}
\label{tm_dis}
\end{figure}

\begin{figure}
\epsfxsize=7.5cm
\epsffile{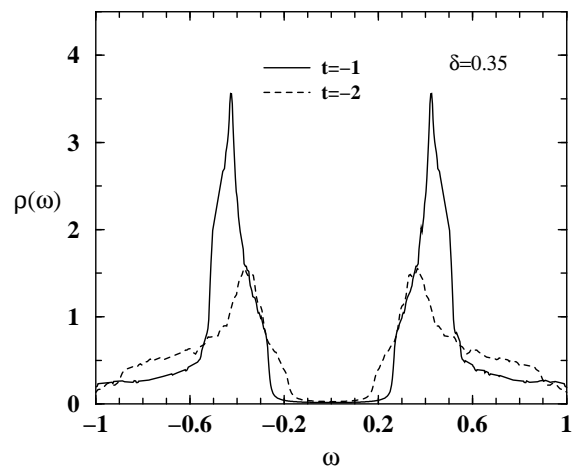}
\caption{The quasi-particle density of states.}
\label{tm_dos}
\end{figure}

Both the quasi-particle dispersion and the density of states, in Figures~\ref{tm_dis} and~\ref{tm_dos} respectively,
 show an energy gap in the spectrum.  The reason lies in the fact that function D({\bf k}) is complex. For
$(\theta,\phi)=(2\pi/3,4\pi/3)$, the function $\mbox{D}({\bf
k})=d_1-id_2$, where $d_1 = \cos(k_x) - \cos(k_x/2)\cos(\sqrt{3}k_y/2)$ and $d_2 =
\sqrt{3}\sin(k_x/2)\sin(\sqrt{3}k_y/2)$. Thus our results are akin to the $d_{x^2-y^2}+ i d_{x y}$ symmetry 
case in the cuprates.

\begin{figure}
\epsfxsize=7.5cm
\epsffile{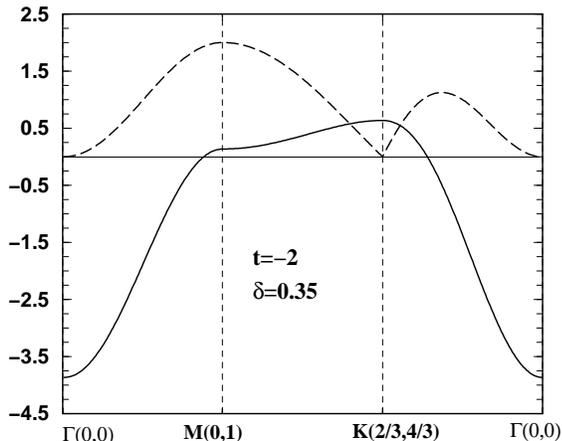}
\caption{The  dashed  line is for the $|$D({\bf k})$|$ and the solid line is for $(\epsilon({\bf k})-\mu)/2t\delta$.}
\label{epD}
\end{figure}

The modulus of D({\bf k}) is non-zero at all points in the Brillouin zone, except at six zone corner points, that is 
$(2\pi/3,-2\pi/3)$, $(2\pi/3,4\pi/3)$ etc  and the origin. Though $\epsilon({\bf k})-\mu$ has contour of zeros within the 
Brillouin zone, the chance of six corner points lying on this contour of zeros is almost zero. This point is clear in
Figure~\ref{epD}, where $|$D({\bf k})$|$ and $\epsilon({\bf k})-\mu$ are plotted along three symmetry directions of
the Brillouin zone. 
The gap for $t=-1$ and $-2$ is approximately 0.2J
and 0.1J, respectively.

The gap ($\sim |\Delta|$)  decreases with increasing $|t|$ as already noted. We   mention that
the quasi-particle spectrum is gapped for $t>0$ as well, but  the gap is  very small around the doping of our interest. In
fact for reasonable values of $t$ (say, 3) it is zero, precisely because $\Delta=0$. Thus, Case B does not favor RVB
solution for large dopings, say 0.3 or beyond.

\begin{figure}
\epsfxsize=7.5cm
\epsffile{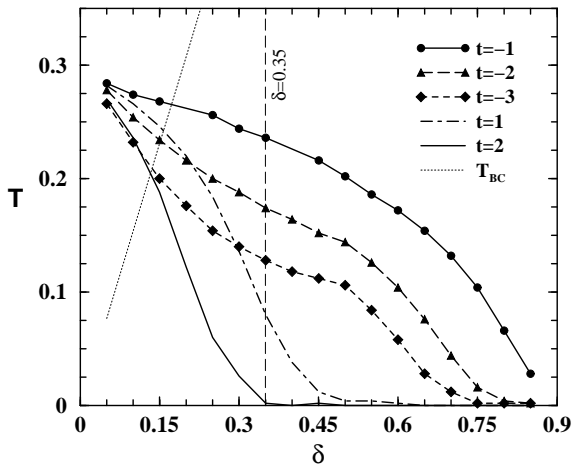}
\caption{T is in units of J. The case with $t>0$ for  $\delta \sim 0.35$  is not very robust for the RVB
         mean field theory whereas  $t<0$ is favourable.}
\label{d_T}
\end{figure}

We have also calculated temperature at which the mean field order parameter  $\Delta$ vanishes for different values 
of $\delta$. It helps us understand the broad nature of thermodynamic phase diagram in T-$\delta$ plane.
Figure~\ref{d_T} shows the transition temperatures for different hole concentrations for which $\Delta$ vanishes.

The question of RVB superconductivity needs  a little more care than what we have given so far. The mean field $\Delta$,
though a pairing order parameter, doesn't by itself  imply  superconductivity. For example at half
filling $\Delta$ is non-zero, but it is insulating.
The identification of the superconducting phase in T-$\delta$ diagram can, however, be done within the
frame work of the slave boson approach. This approach has been quite popular in the RVB theories of t-J model on
square lattice~\cite{ubbens_lee}. Here  the physical electron operator, $c^\dag_\sigma = f^\dag_\sigma b$,
where $f$ is the spin-1/2 neutral fermion, and $b$ is the spin-0 charged boson, and the projection of double occupancy
is in-built into the construction.

The superconducting order parameter $\langle c^\dag c^\dag\rangle \sim 
\langle b b\rangle\langle f^\dag f^\dag\rangle$  is a product of spin pairing order parameter and a bose condensation
factor. What we have got from the mean field calculation is essentially the spin-pairing order parameter. The true
superconducting order parameter is $\Delta_{SC} \sim F_B\Delta$, where $\Delta$ is the mean field order parameter. The 
bose condensation temperature for $b$ bosons (for $F_B$ to be non-zero) needs to be estimated separately.
That region of T-$\delta$  diagram, where
both $F_B$ as well $\Delta$ are non-zero, can be interpreted as the RVB superconducting phase. There are three more
typical regions according to this interpretation : (1) Spin-gap : $\Delta\neq 0$ and $F_B=0$, (2) Strange metal :
$\Delta = F_B = 0$, and (3) Normal metal : $\Delta =0$ and $F_B\neq 0$. With this qualitative picture in mind, we will
present a rough phase diagram for the cobalt oxide superconductors.

Let us briefly mention how we estimate the bose condensation temperature, T$_{BC}$, for the bosons. There
are other ways~\cite{ubbens_lee}, but we discuss our roughly equivalent prescription.
 The $b$ bosons, at the level of mean field
decoupling, will effectively have the same band structure as that of the tight binding electrons on triangular
lattice. Therefore, the T$_{BC}$, is defined as a temperature at which the boson chemical potential tends to be the
band's bottom. Since free bosons can not condense in two dimensions, we will take the case of three
dimensional band structure with a large c-axis anisotropy. Around the band's bottom, the energy dispersion can be 
approximated as $c^*(k^2 + k_z^2/\gamma)$. The curvature, $c^*$, is related to the two dimensional density of states 
at the band edge,
$\rho^*$ as : $c^*\approx \frac{1}{4\pi\rho^*}$. Now, using the fact that free boson can condense in three dimensions,
\(\delta = \frac{1}{L}\sum_{\bf k}1/(e^{\beta_{BC}(\epsilon({\bf k})-\epsilon_{min})}-1)\).
Simplifying it for $\gamma\gg\pi/4$, we get
\[T_{BC}\approx \frac{\delta}{\rho^*}\frac{1}{2+\log(4\gamma/\pi)}\]
Since the anisotropy affects only logarithmically, we can safely take some large value for $\gamma \sim 100$, especially when
the real system is a good quasi two dimensional system.

\subsection{Phase Diagram}
\begin{figure}
\epsfxsize=7.5cm
\epsffile{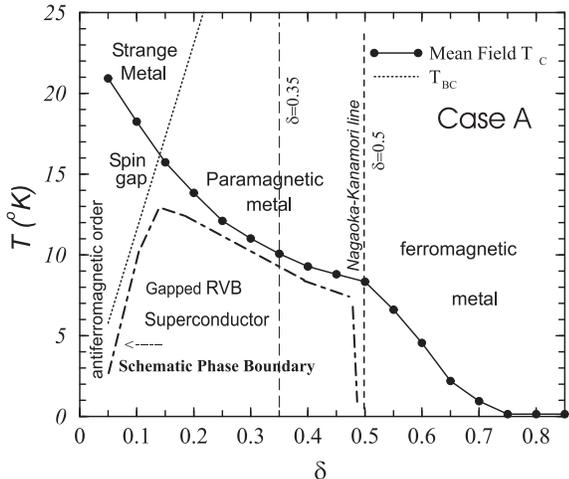}
\caption{The mean field phase diagram for $t=-3$J and ($\theta,\phi$)$=$($2\pi/3,4\pi/3$).}
\label{phase_diag_tm3}
\end{figure}

Here, we propose a rough phase diagram for triangular lattice, layered cobalt oxide materials over a wide range of 
doping based upon our  calculation in the scheme of Case A(ii) as  shown in Figure~\ref{phase_diag_tm3}. This phase diagram  encompasses a number of interesting phases, including the recently observed superconducting phase. 
We will briefly describe each of the labelled region in Figure~\ref{phase_diag_tm3}. 
The  exchange coupling J is estimated from the high temperature susceptibility data~\cite{ray} for Na$_{0.5}$CoO$_2$. 
Its Curie Weiss temperature is approximately $-118$K which gives J$\sim 79$K. 
For doping $\delta=0.35$, the T$_C$ for transition into an RVB superconducting state is approximately 10K, 
compared to  the measured T$_C \sim 5$K~\cite{takada}.
An important prediction for the superconducting phase is the
existence of an energy gap in the quasi-particle spectrum. It arises because the superconducting
order parameter is complex, with relative internal phases $2\pi/3$ and $4\pi/3$ thus avoiding the
possibility of generic vanishing of the gap along lines or points in the momentum space.
 The optimal doping seems to be around 0.15.

On the higher side of the doping $ \delta \gtrsim 0.5$ in Figure~\ref{phase_diag_tm3},
 we expect the RVB superconducting state to become
unstable in favor of a ferromagnetic metallic state where either the Nagaoka or the Kanamori ferromagnetism plays a
dominant role. We call $\delta=0.5$ (electron concentration, $n=1.5$) as the Nagaoka-Kanamori line,
as  the rough location where the Nagaoka physics of the doped Mott state transmutes  into the
Brueckner type Kanamori physics of multiple scattering in an almost filled band.
The precise location of the phase boundary between the superconducting  and the ferromagnetic state
is difficult to predict from the present calculation, since the RVB mean field theory is not particularly 
accurate in getting the absolute numbers for various states. We present  only a guide to the eye in our phase diagram.
At sufficiently high temperatures, we expect ferro-metal to para-metal transition across the Nagaoka-Kanamori line
which can be estimated from the spin stiffness of the quasi 2-d ferromagnet.  
Apart from the above mentioned important phases, one may expect spin-gapped phase in the low doping, low temperature
regime, since that is where exchange interaction will play more decisive role. The  three sublattice  antiferromagnetic phase  of the triangular lattice AFM  is likely to persist for small doping. We also expect a strange metallic phase to exist at moderate doping, but  only at higher 
temperatures.

\begin{figure}
\epsfxsize=7.5cm
\epsffile{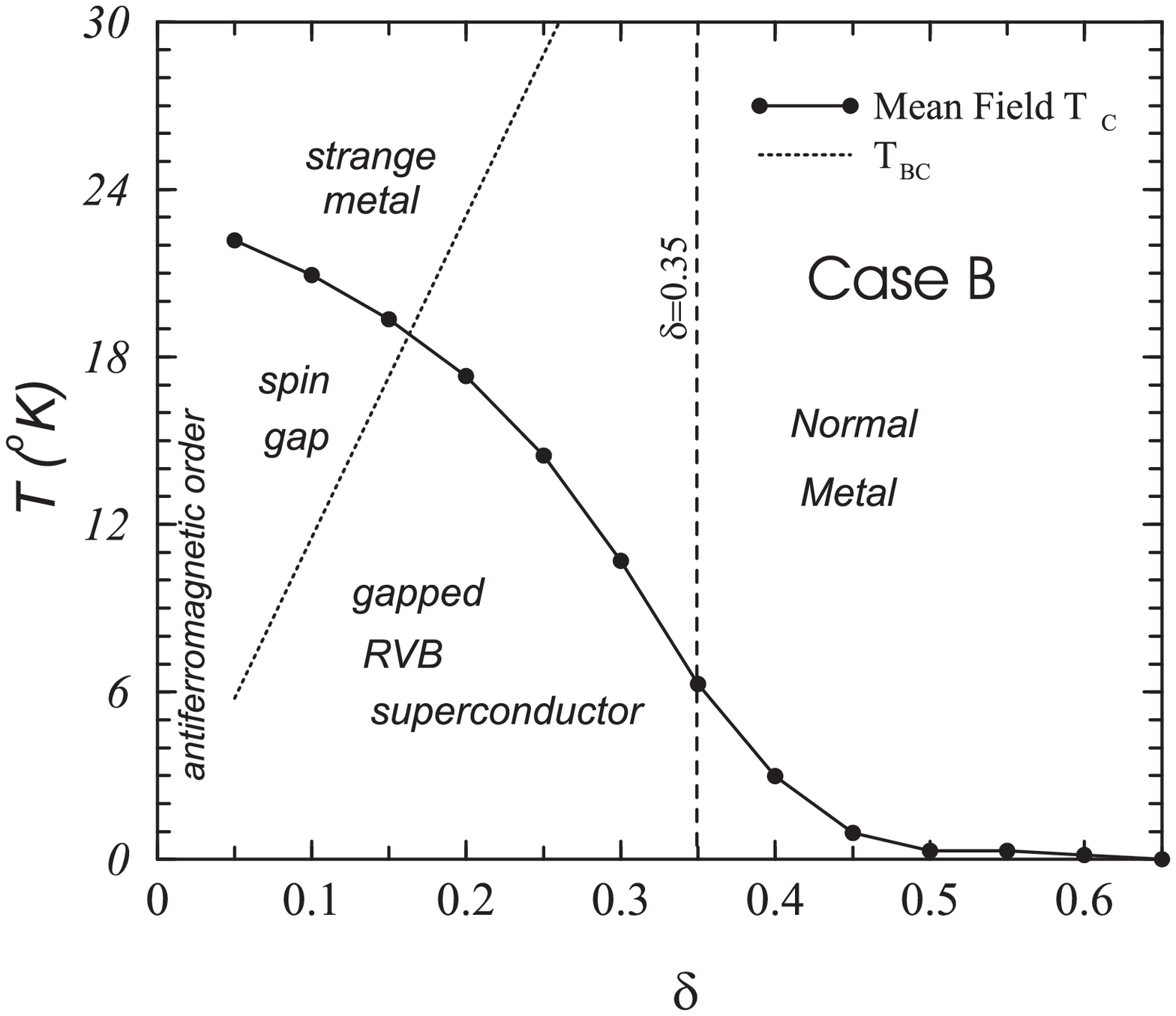}
\caption{The mean field phase diagram for $t=\mbox{J}$ and ($\theta,\phi$)$=$($2\pi/3,4\pi/3$).}
\label{phase_diag_tp1}
\end{figure}

We next also present, in
Figure~\ref{phase_diag_tp1}, the mean field phase diagram for Case B(i). Though Case B is less favourable
 than Case A,  it does  yet achieve the RVB superconducting phase for small values of $t$ ($\sim\mbox{J}$ or smaller; see Figure~\ref{d_T}). 
Since the value of $|t|$ in these cobaltates is believed to be rather low, it seems appropriate to record the
results for Case B also.
This phase diagram  is a simpler version of the
mean field phase diagram for Case A(ii), with the symbols having the same meaning.
The possibility of a ferromagnetic phase does not arise in this case, leading to its
simplicity.  Here we find   T$_C\sim 6$K at 0.35 hole concentration. The 
optimal doping and the maximum T$_C$ are similar to those of Case A, but  T$_C$ falls off
more rapidly for $\delta \geq 0.15$.

\section{Summary}
To summarize the work presented here, we have performed an RVB mean field theory on the triangular lattice  inspired by the
recent discovery of the superconductivity in cobalt oxide layered materials. 

Our use of a single band model in the present context needs a word of justification.
The LDA band structure for Na$_{0.5}$CoO$_2$ \cite{dsingh,comment} indicates the possibility of 
a competition between two bands for the fermi level on moving away from that doping in either direction.
 One would conclude that $\delta_c \sim 0.5$ if the bands are held rigidly  under doping, since at this
filling the fermi level tangents the  (filled) band emanating from the  $\Gamma$ point in Fig.1 of Reference~\cite{dsingh}.
The systems at different dopings would have their fermi levels 
in different bands.
 These would then correspond to 
(1) Case A(i) for $\delta < \delta_c$ i.e.   electron like   and  (2) B(ii) for $\delta > \delta_c$, i.e. hole like. 
A  ``one band model''
appears to be justifiable provided one is sufficiently far from $\delta_c$.
In view of the uncertainty in the exact location of $\delta_c$, we have presented the results for both signs of t above. 
While photoemission may shed light on this issue, its interpretation needs caution in view of
 the remarkable possibility  that
the observed renormalized  fermi surface in strongly correlated systems could differ  drastically 
in shape from the {\it bare} fermi surface, that has emerged from  recent numerical work \cite{himela_ogata,puttika}.
Our concern in this work is with the ``bare fermi surface'', i.e. with the sign of the {\it bare} $t$.
 
We argue that the superconducting material $\delta=0.35$
corresponds to an effective one band case with $t>0$ and electron doping, i.e. A(i). This is logical
since a filling of $\delta=0.35$ seems far enough from any reasonable $\delta_c$.
Several detailed results are presented.
We find the symmetry of the superconducting order parameter, and predict the existence of an energy gap. We also present an approximate phase diagram
for the cobalt oxide superconductor in the T-$\delta$ plane.

In the more interesting Case A, there is a competition  between ferromagnetic order and superconductivity.
Our  low temperature RVB superconducting phase basically arises in the doping range where
Nagaoka ferromagnetism is suppressed by the antiferromagnetic J. The superconductivity found here
  may thus be viewed
as arising  from  what may be termed {\em total frustration} , i.e. from the
competition  between
 competing terms,   the  frustrated electronic motion that  weakly prefers Nagaoka  ferromagnetism, and  the frustrated spin exchange that prefers either no long ranged order ( i.e.  a short ranged RVB  type state)  or a doping weakened   three sublattice order.  
 
One major assumption in this work is that the scale of $t$ is not too different from J which is estimated to be
around $79$K. This estimate is at odds with the LDA estimate of the band width by two orders of magnitude, and thus we 
cannot claim to have ``explained'' the low degeneracy temperature scale, we have merely assumed it and worked out 
the consequences for other properties. Indeed the emergence of a low energy scale in these systems seems to be a 
central problem of the cobaltates.

Finally we have presented our results for the  high frequency Hall constant $R_H^*$.  This is predicted to grow linearly with
temperature without saturation, and the slope depends in a known way on $t$ and the filling.  
The filling dependence $ \frac{1+\delta}{\delta (1- \delta)}$ should be readily testable.
The explicit expressions enable one to extract the hopping matrix element $t$. 
The unusual  behaviour of the Hall constant
 is a special property of the triangular lattice, having to do with its unique topology of smallest
length of closed loops, with odd length. Our results are valid for  high temperatures ( $k_B\mbox{T}\gg |t|$ ).
Remarkably enough, this condition appears easy to fulfill in the cobaltates, and hence it should be possible
to utilize these results to extract basic parameters for the system from high frequency Hall experiments.
Our result might also  be useful in interpreting transport Hall data.

BK  acknowledges  CSIR for financial support.
BSS was supported in part by an Indo French grant IFCPAR/2404.1

\end{multicols}

\begin{references}
\bibitem{comment} Our assignment of the sign of ``$t$'' leads to an electron like fermi surface surrounding the $\Gamma$ 
                  point, which might be ascribable to the electron like band in Figure 1 of Reference~\cite{dsingh} 
                  having a minimum at the $\Gamma$ point. This is reasonable if remember that the electron count for 
                  the superconducting samples $x=0.35$ forces the fermi level to dip below the value for the $x=0.5$ case 
                  reported in Reference~\cite{dsingh} which has a hole like fermi surface surrounding the $\Gamma$ point.
\bibitem{note1} We get lattice parameters $a=2.823$\AA~and $c=19.621$\AA~from Reference~\cite{takada}. The physical
                cell volume containing one cobalt ion, $v = \frac{1}{2}\frac{\sqrt{3}a^2c}{2}$. The factor of 1/2
                is there because each crystallographic unit cell of Na$_x$CoO$_2\cdot y$H$_2$O contains two CoO$_2$ 
                layers, each contributing one cobalt ion.
\bibitem{note2} With an absolute scale measurement of $R_H^*$, it should be possible to obtain not only the sign of
                ``$t$'', but also its magnitude. Optical Hall constant measurements may be roughly guessed at from the 
                transport measurements.
\end{references}
\end{document}


\title{Erratum: Superconductivity in CoO$_2$ Layers and the  Resonating Valence Bond Mean Field Theory of the Triangular 
       Lattice t-J Model\\
{[Phys. Rev. {\bf B 68 }, 104508  (2003)]}}
\author{Brijesh Kumar$^{1,2}$ and B. Sriram Shastry$^{1,3}$ }
\address{(1)Department of Physics, Indian Institute of Science, Bangalore 560012, India.\\
(2) ITP, \'{E}cole Polytechnique F\'{e}d\'{e}rale de Lausanne, CH 1015 Lausanne
Switzerland\\
(3) Department of Physics, University of California Santa Cruz, Ca 95064 }
\maketitle

\vspace{5mm}

Our recent paper \cite{ks} contains an error in the formula for the high frequency Hall constant for the 
experimentally relevant case of electron doping.  Eq(3) giving the   high frequency  Hall constant for the case of electron filling, i.e.  electron density $ n > 1$, should correctly  read  
$$
R_H^* =   -  \frac{v}{4 |e|} \frac{k_B T}{ t}\frac{1+\delta}{\delta (1- \delta)} \hspace{2cm} \mbox{(Cases A(i) or B(ii))}  \hspace{2cm}\mbox{Eq(3')}
$$
where $\delta = |1-n|$.  This differs essentially from the quoted answer  by a minus sign that was missed by us,
and by a factor of 2. This equation is  obtained from Eq(2) in the paper,  calculated  for the case of hole doping $n < 1$, by a particle hole transformation.    Performing a   canonical 
transformation $c_{i,\sigma} \leftrightarrow c^\dagger_{i,\sigma}$
in  the Kubo formulas for models of lattice fermions (see e.g. \cite{sss} Eqn(1,2) )   yields:
$$ 
R_H(n , t ,\omega,T) = - R_H(2- n, -t, \omega, T).
$$
The crucial overall minus sign  arises from the fact that the Hall constant involves a three current propagator, and each current picks up 
a minus sign under the transformation. Incidentally this identity results in a vanishing at half filling for  bipartite lattices of the Hall function $R_H(1,t,\omega,T)$ for any value of $T,\omega$ and the interaction strength.

The changed sign  in Eq(3') has  a few implications  which we would like to  state explicitly. 
The recent experiment of Wang, Rogado, Cava and Ong\cite{ong} on the  transport Hall constant, actually sees the
 predicted linear T behaviour    over a wide temperature range $ 200^0 K \leq T \leq 400^0K$ with a positive slope.
From Eq(3') above,  we deduce  that the appropriate $t-J$ model description of the system has
$t<0$,  i.e. belongs to  case B(ii) in our notation, at least at the composition $x\sim .7$. 

We note that with this correction, the sign of the hopping $t$ suggested by our calculation {\em is the same} as the one 
favoured by Refs\cite{baskaran,lee}, at least for the non superconducting range  $x\sim .7$. In this  composition range, photoemission data \cite{photoemission1,photoemission2} also suggest the same sign of $t$, with an {\em unoccupied region} around the center of Brillouin zone, namely the  $\Gamma$ point.
